\documentclass{article}
\usepackage{sp_conf}

\usepackage{graphics}

\begin{document}

\name{Jean-Marc Valin and Roch Lefebvre\thanks{This work was financed by VoiceAge Corp. and the NSERC. E-mail: jmvalin@jmvalin.ca}\thanks{\copyright 2000 IEEE.  Personal use of this material is permitted. Permission from IEEE must be obtained for all other uses, in any current or future media, including reprinting/republishing this material for advertising or promotional purposes, creating new collective works, for resale or redistribution to servers or lists, or reuse of any copyrighted component of this work in other works.}}

\address{University of Sherbrooke, Department of Electrical Engineering\\Sherbrooke, Qu\'ebec, J1K 2R1, Canada}

\ninept

\title{Bandwidth Extension of Narrowband Speech for Low Bit-Rate Wideband Coding}

\maketitle
\begin{abstract}
Wireless telephone speech is usually limited to the \( 300-3400\, \mbox{Hz} \) band,
which reduces its quality. There is thus a growing demand for wideband speech
systems that transmit from \( 50\, \mbox{Hz} \) to \( 8000\, \mbox{Hz} \). This paper presents
an algorithm to generate wideband speech from narrowband speech using as low
as \( 500\, \mbox{bits/s} \) of side information. The \( 50-300\, \mbox{Hz} \) band is predicted
from the narrowband signal. A source-excitation model is used for the \( 3400-8000\, \mbox{Hz} \)
band, where the excitation is extrapolated at the receiver, and the spectral
envelope is transmitted. Though some artifacts are present, the resulting wideband
speech has enhanced quality compared to narrowband speech.
\end{abstract}

\section{Introduction}

Communication systems increasingly use wideband speech in applications such
as videoconferencing and teleconferencing. However, the PSTN and wireless networks
mostly use the \( 300-3400\, \mbox{Hz} \) telephone band. Part of this limitation
is due to the \( 8\, \mbox{k\mbox{Hz}} \) sampling rate used in those systems. One way to
increase the audio bandwidth is to increase the sampling frequency to \( 16\, \mbox{k\mbox{Hz}} \).
However, this is not practical in many situations either because of bandwidth
limitations (for example in wireless systems) or because it would require huge
modifications to legacy systems (as in the PSTN). Another approach is to use
the spectral redundancies of speech to recover the wideband components from
the received narrowband speech.

We propose, in this paper, a system to recover wideband speech from narrowband
speech using a small amount of side information. Unlike other approaches \cite{codebook, MFCC},
which attempt to recover the \( 4000-8000\, \mbox{Hz} \) band from the \( 0-4000\, \mbox{Hz} \)
band, the proposed system recovers both the low-frequency (\( 50-300\, \mbox{Hz} \))
and the high-frequency (\( 3400-8000\, \mbox{Hz} \)) bands using only the \( 300-3400\, \mbox{Hz} \)
telephone band and additional side information. The objective is to minimize
the bit-rate of this side information towards an ultimate goal of \( 0\, \mbox{bit/s} \).
The system uses a speech-specific model, hence it does not work well for general
audio.

\section{System overview}

\begin{figure}[t]
{\centering \resizebox*{1\columnwidth}{!}{\includegraphics{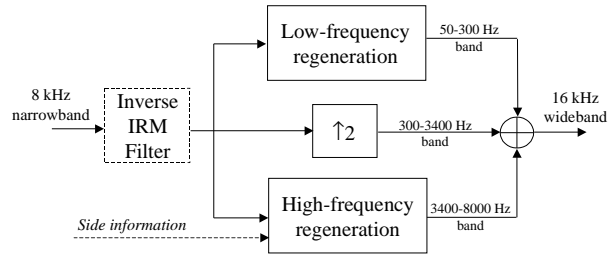}} \par}

\caption{Overall system\label{system}}
\end{figure}

The complete system block diagram is shown in Figure \ref{system}. The inverse
IRM filter is an FIR filter which, once convolved with the IRM filter, gives
a flat response in the \( 200-3500\, \mbox{Hz} \) range. Since not all voice systems
use IRM filtered speech, this inverse filter is optional. For the remaining
of the paper, the term ``narrowband'' signal will refer to the output of the
inverse IRM filter. Note that the low-frequency and high-frequency bands are
regenerated using different approaches. It is also worth noting that the up-sampling
block includes the necessary anti-aliasing filter.

\section{Low frequency regeneration}

\begin{figure}[t]
{\centering \resizebox*{1\columnwidth}{!}{\includegraphics{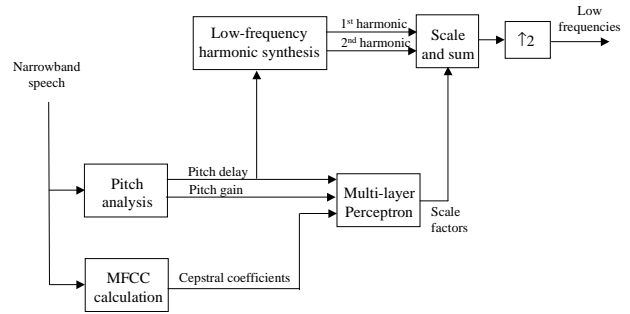}} \par}

\caption{Low-frequency regeneration scheme\label{fig_LF}}
\end{figure}

For the low frequencies (\( 50-300\, \mbox{Hz} \)), we make the assumption that in
this band, voiced speech can be represented as a set of, at most, two sinusoids
(i.e. the first two harmonics). A controlled sinusoidal oscillator is used to
generate the pitch harmonics. The frequency of the oscillator is given by the
pitch analysis and the phase is adjusted so that it remains coherent across
frames. We make the assumption that the absolute phase is not too perceptually
relevent.

\subsection{Harmonic amplitudes prediction}

Once the two harmonics are generated in the lower band, they must be scaled
to the right amplitude (Figure \ref{fig_LF}). The scaling factor is estimated
using a multi-layer perceptron network. The network we use has 18 inputs:

\begin{itemize}
\item 16 mel frequency cepstral coefficients calculated on the narrowband signal,
and
\item the pitch gain and delay.
\end{itemize}
The two outputs of the network are the gains (in the log domain) for each of
the two synthesized harmonics. We use 2 hidden layers with 10 units per layer.
The two hidden layers use a hyperbolic tangent as the activation function, while
the output layer uses linear activation functions, so that the outputs are not
restricted to \( [-1,1] \) interval.

\subsection{Harmonic amplitude evaluation}

In order to evaluate the harmonics amplitude for training the neural network,
the first two harmonics are extracted using a least-square fit of windowed sinusoids.
The amplitude of the two sinusoids is calculated as

\begin{equation}
\label{eq_LS}
\mathbf{a}=\left( \mathbf{X}^{T}\mathbf{X}\right) ^{-1}\mathbf{X}^{T}\mathbf{y}
\end{equation}
 where \( y \) is the windowed, rectified narrowband signal and \( X \) is
a 128 by 5 matrix where each column is one of the basis functions for the current
frame. The first basis function is a Hanning window (windowed DC), while the
others are windowed sines and cosines at the first and second harmonic. For
instance, the second basis function is
\begin{equation}
\label{eq_LS2}
x_{2}(n)=\left[ .5-.5\cos \left( \frac{2\pi n}{L}\right) \right] \sin \left( \frac{2\pi n}{T}\right) 
\end{equation}
where \( L \) is the length of the window and \( T \) is the pitch period.
The accuracy of the process depends on pitch evaluation. Some of the problems
encountered are due to pitch period doubling. The problem seems to be worse
for females than males.

\section{High frequency regeneration}

\begin{figure}[t]
{\centering \resizebox*{1\columnwidth}{!}{\includegraphics{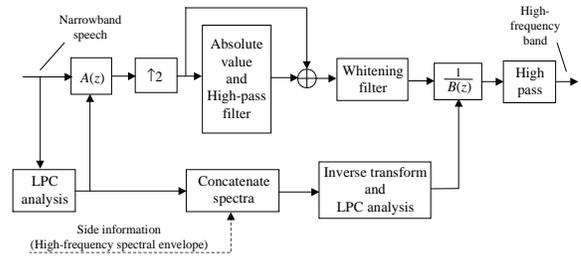}} \par}

\caption{High frequency regeneration scheme\label{fig_HF}}
\end{figure}

For high frequency reconstruction, we use an LPC analysis to divide the signal
into a spectrally-flat excitation and a synthesis filter representing the spectral
envelope. The excitation and the envelope are extended independently in the
\( 3400-8000\, \mbox{Hz} \) band, as shown in Figure \ref{fig_HF}. It is important
to note that the input of the process is pre-filtered to emphasize high frequencies.
This improves the LPC analysis accuracy.

\subsection{Excitation extension}

Because of the phase considerations in extending the excitation, it is much
easier to work in the time domain. There are two widely used methods to perform
excitation extension in the time domain \cite{OLD}:

\begin{itemize}
\item Over-sampling excitation, which causes spectrum folding by aliasing, and
\item Using a non-linear function to generate harmonics at higher frequencies.
\end{itemize}
Because of the ``artificial tones'' caused by the folding me\-thod, we prefer
the non-linearity approach. The absolute value function is a good candidate
for this method since, unlike the square value, it does not require energy normalization.
The wideband excitation generated in this fashion is phase-coherent with the
original narrowband excitation and preserves the harmonic structure without
any discontinuity in the spectrum. Unlike the spectral folding technique which
produces very tonal harmonics at higher frequencies, the absolute value function
increases the noise level between the high-frequency harmonics (sounding more
natural). 

The whitening filter is used to flatten the spectrum of the extended excitation
so that it is similar to the excitation that would have been calculated from
wideband speech. This is done by performing an LPC analysis and using the resulting
coefficients to whiten the extended excitation. A gain must then be applied
so that the low-frequency energy is equal to the low-frequency energy of the
narrowband excitation.

\subsection{Spectral envelope coding}

Several approaches were considered to extrapolate the high-frequency spectral
envelope from the narrowband spectral envelope. In all cases, the subjective
quality was not satisfactory. This suggests that the high-frequency formant
structure of speech cannot be accurately predicted from the narrowband formants.
Hence, side information is used to transmit the high-frequency spectral envelope
as shown in Figure \ref{fig_HF}.

This spectral information is transmitted in the transform domain. Specifically,
the power spectrum of an LPC filter is first calculated on the original wideband
speech. After proper energy scaling, the \( 3400-8000\, \mbox{Hz} \) band of this
power spectrum is vector quantized in the log-domain using 40 frequency-points
(64 points for the whole \( 0-8\, \mbox{k\mbox{Hz}} \) spectrum). The receiver concatenates
this high-frequency spectral information with its local calculation of the narrowband
spectral envelope (Figure \ref{fig_HF}). The full-band LPC filter (\( 1/B(z) \)
in Figure \ref{fig_HF}) can then be recovered by an inverse transform followed
by Levinson-Durbin recursion.

It was found that an 8-bit vector quantizer was sufficient to transmit a good
estimate of the high band. We use a VQ trained with a variant of the LBG algorithm.
With a frame size of 256 samples at a sampling rate of \( 16\, \mbox{k\mbox{Hz}} \), the
bit rate necessary to transmit this side information is minimal --- \( 500\, \mbox{bits/s} \).
The objective of this work is to reduce this side information rate as close
to zero as possible.

\section{Results}

Because the phase of the original wideband signal is not preserved at both low
and high frequencies, measuring the system performance using objective measures,
such as the SNR, is not possible. However, it has been noted that for the high-frequency
band, the effect of replacing the original wideband excitation by the reconstructed
excitation is very small. This leaves the envelope reconstruction as the main
cause of distortion. Using a codebook of 256 vectors for the high-frequency
envelope, we obtain a mean spectral distortion of \( 3.6\, \mbox{dB} \) over the \( 3-8\, \mbox{k\mbox{Hz}} \)
range. This spectral distortion could be further reduced by coding the difference
between the real envelope and a prediction. Note that the training set and test
set are disjoint. Each contains 40 minutes of speech after removing all silence. 

The result of the highband recovery is perceptually very close to the original
wideband speech. However, unlike the high-frequency problem, the error in the
estimation of the low-frequency harmonics is perceptible. Some artifacts are
caused by the fact that the amplitude of the added sinusoids are not coded.
The mean error in estimating the low-frequency harmonics amplitudes with a multi-layer
perceptron network is \( 3.0\, \mbox{dB} \).

When starting from \emph{coded} narrowband speech using either G. 729 (at \( 8\, k\mbox{bits/s} \))
or EFR GSM (at \( 13\, k\mbox{bits/s} \)), we obtain a quality comparable to the case
when the narrowband speech is not coded.

\begin{figure}[t]
{\centering \resizebox*{1\columnwidth}{!}{\includegraphics{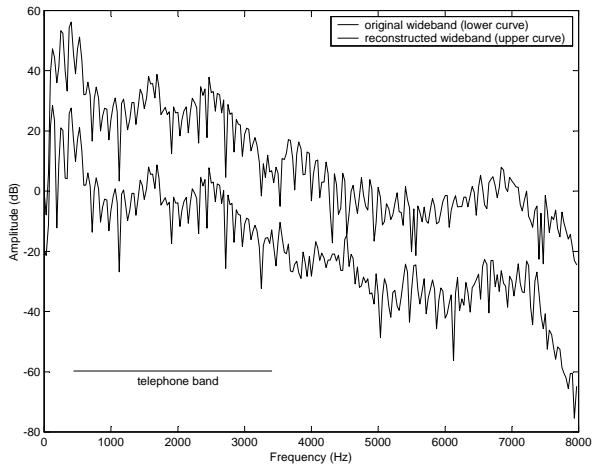}} \par}

\caption{Wideband reconstruction for voiced speech\label{fig_voi}}
\end{figure}

\begin{figure}[t]
{\centering \resizebox*{1\columnwidth}{!}{\includegraphics{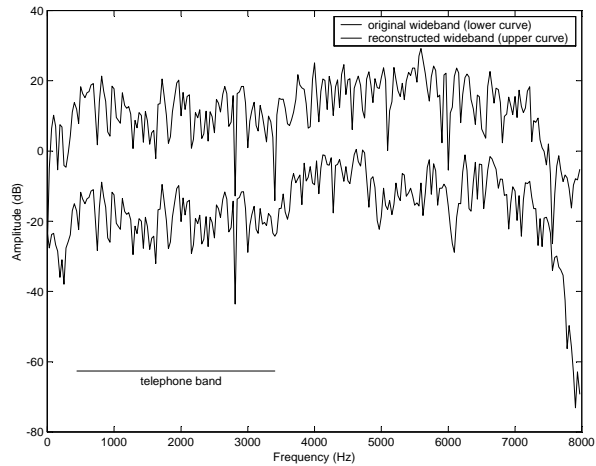}} \par}

\caption{Wideband reconstruction for unvoiced speech\label{fig_unv}}
\end{figure}

Figures \ref{fig_voi} and \ref{fig_unv} show the reconstructed wideband (with
a \( 30\, \mbox{dB} \) offset) compared to the original wideband for both voiced and
unvoiced speech. The spectra were calculated using a \( 32\, ms \) Hanning
window.

\section{Conclusion}

In this paper, we proposed a method to recover wideband speech using narrowband
speech and minimal side information. The resulting wideband speech has enhanced
quality compared to narrowband speech. Although the high-frequency band can
be almost perfectly reconstructed, the low-frequency band still contains some
noticeable artifacts.

The proposed system could be used to transform a narrowband codec into a wideband
codec by adding as little as \( 500\, \mbox{bits/s} \) to the transmitted information.

\end{document}